\documentclass[journal]{IEEEtran}

\ifCLASSINFOpdf
\else
\fi
\usepackage{cite}
\usepackage{amsmath,amssymb,amsfonts}
\usepackage{algorithm}
\usepackage{algorithmic}
\usepackage{graphicx}
\usepackage{textcomp}
\usepackage{xcolor}
\usepackage{stfloats}
\usepackage{caption}
\usepackage[T1]{fontenc}
\usepackage{array}
\usepackage{booktabs}
\newtheorem{remark}{Remark}

\begin{document}
%
\title{Anti-Intercept OFDM Waveform Design with Secure Coding for Satellite Networks}

\author{Zhisheng~Yin,~\IEEEmembership{Member~IEEE,}
        Yonghong~Liu,~\IEEEmembership{Student Member~IEEE,}
        Dongbo~Li,~\IEEEmembership{Member~IEEE,}
        Nan~Cheng,~\IEEEmembership{Senior Member~IEEE,}
        Linlin~Liang,~\IEEEmembership{Member~IEEE,}
        Changle~Li,~\IEEEmembership{Senior Member~IEEE,}
        ~and Jie~Liu,~\IEEEmembership{Fellow~IEEE}
\thanks{
This work was supported in part by the National Natural Science Foundation of China under Grant 62201432, Grant 62071356, Grant 62101429, and Grant No.62350710797, and in part by the Innovation Project of the National Key Laboratory of Smart Farm Technologies and Systems under Grant No. JD2023GJ01.

Corresponding author: Yonghong Liu.

Zhisheng Yin, Nan Cheng, and Changle Li are with the State Key Laboratory of ISN and the School of Telecommunications Engineering, Xidian University, Xi’an 710071, China (e-mail: zsyin@xidian.edu.cn; dr.nan.cheng@ieee.org; clli@mail.xidian.edu.cn.)

Yonghong Liu is with the State Key Laboratory of ISN and the School of Cyber Engineering, Xidian University, Xi’an 710071, China (e-mail: 23151214131@stu.xidian.edu.cn.)

Dongbo Li is with the Faculty of Computing and the National Key Laboratory of Smart Farm Technologies and Systems,
Harbin Institute of Technology, Harbin 150001, China (e-mail: ldb@hit.edu.cn)

Linlin Liang is with the School of Cyber Engineering, Xidian University, Xi'an 710071, China (llliang@xidian.edu.cn)

Jie Liu is with the International Research Institute for Artificial Intelligence, Harbin Institute of Technology, Shenzhen 518055, China, and also with the National Key Laboratory of Smart Farm Technology and Systems, Harbin Institute of Technology, Harbin 150001, China (e-mail: jieliu@hit.edu.cn)
}
}
\markboth{Journal of \LaTeX\ Class Files,~Vol.~14, No.~8, August~2015}%
{Shell \MakeLowercase{\textit{et al.}}: Bare Demo of IEEEtran.cls for IEEE Journals}

\maketitle

\begin{abstract}
Low Earth Orbit (LEO) satellite networks are integral to next-generation communication systems, providing global coverage, low latency, and minimal signal loss. However, their unique characteristics, such as constrained onboard resources, Line-of-Sight (LoS) propagation, and vulnerability to eavesdropping over wide coverage areas, present significant challenges to physical layer security. To address these challenges, this paper focuses on the design of anti-intercept waveforms for satellite-ground links within Orthogonal Frequency Division Multiplexing (OFDM) systems, aiming to enhance security against eavesdropping threats. We formulate a secrecy rate maximization problem that aims to balance secrecy performance and communication reliability under eavesdropping constraints and sub-carrier power limitations. To solve this non-convex optimization problem, we propose a bisection search-activated neural network (BSA-Net) that integrates unsupervised learning for secure coding optimization and bisection search for dynamic power allocation. The proposed method is structured in two stages: the first optimizes secure coding under power constraints, while the second allocates power across sub-carriers under eavesdropping constraints. Extensive simulation results demonstrate the efficacy of our approach, showcasing significant improvements in secrecy rate performance.
\end{abstract}

\IEEEpeerreviewmaketitle
\section{Introduction}

\IEEEPARstart{W}{ith} the gradual commercialization of the fifth-generation (5G) mobile network, research on the sixth-generation (6G) mobile communication has also commenced. Compared to 5G, the 6G standard demands higher communication quality and broader coverage \cite{b1}. LEO satellites have garnered renewed academic interest due to their advantages, including extensive coverage, low latency, minimal signal attenuation, high resilience, and uninterrupted all-weather operation \cite{b2b}, \cite{b2}. Countries and enterprises are actively constructing LEO satellite communication systems \cite{b3}, including initiatives such as Starlink, Iridium, Globalstar, Orbcomm, and ARGOS. LEO satellite systems widely adopt orthogonal frequency division multiplexing (OFDM), a multiplexing technique introduced by Chang et al. in 1966 \cite{b4}, due to its high spectral efficiency and data transmission rates \cite{b5}. Despite these advancements, the inherent vulnerabilities of wireless channels pose critical security risks to satellite-ground communications. The broadcast nature of wireless communications makes satellite-to-ground links especially susceptible to attacks from passive and active adversaries, primarily eavesdroppers (Eves). These adversaries can capture, decode, and recover transmitted signals with sufficient power. 

Consequently, modern communication systems inevitably encounter challenges associated with ensuring communication security. Especially in the past decade, the rapid development of communication technologies has facilitated the transmission of extensive data through wireless networks, thereby amplifying the significance of safeguarding data confidentiality and security. The importance of data transmission security in the communication industry is reflected in industry standards, nearly all of which integrate security algorithms. For instance, LTE employs stream ciphers (e.g., SNOW 3G and ZUC) and block cipher (e.g., AES), whereas GSM relies on the A5 stream cipher \cite{b7}. It is evident that communication system security has always been a critical concern and remains a valuable research topic.

Traditional encryption techniques rely on cryptographic systems based on key management, digital signature authentication, identity verification and related technologies. These security mechanisms are founded on computational cryptography methods, drawing on the design of upper layer protocols in computer networks to ensure information security \cite{b9}. However, the accelerating advancement of communication technologies has posed escalating challenges to traditional key-based systems. First, as the computational power improves and transmission scenarios diversify, the security risks of modern cryptographic encryption algorithms based on computational complexity are becoming more prominent \cite{b10}, \cite{b11}.  In addition, space-ground links require security algorithms to be dynamic because of the diversity, heterogeneity and high frequency of the links \cite{b12}, \cite{b13}. Moreover, the distribution and management of keys face even greater challenges in high eavesdropping-risk environments. Due to the limited satellite resources, these security vulnerabilities become even more critical in space-ground integrated networks. 

Physical layer security is increasingly recognized as a pivotal technique to safeguard sensitive data against eavesdropping threats.This technology exploits the inherent nature of wireless channel, such as channel randomness and reciprocity, to ensure secure and reliable data transmission, eliminating the necessity for complex encryption algorithms \cite{b14}, \cite{b15}. The fundamental principle of physical layer security is to exploits the stochastic and reciprocal nature of wireless channels, enhancing signal quality for authorized users while impairing it for Eves, thereby mitigating the risk of data leakage.
Despite its potential, limited research has been conducted on secure downlink transmission that integrates the unique characteristics of OFDM modulation, particularly in the context of satellite communications. Moreover, existing studies on physical layer security have predominantly focused on key generation, beamforming, and artificial noise injection, with scant attention paid to the design of anti-intercept waveforms. This issue becomes especially critical in satellite communications, where high mobility, dynamic channel conditions, and limited resources collectively pose substantial obstacles to securing downlink transmission within the broader context of wireless security. These pressing concerns underscore the need for innovative solutions, motivating the research presented in this paper, which focuses on addressing the challenges of secure downlink transmission through the design of anti-intercept waveforms in satellite communication systems.

Leveraging the principles of physical layer security, a novel secure coding scheme is proposed for OFDM systems in this paper, specifically designed to enhance anti-intercept capabilities in satellite networks. The proposed scheme exploits the inherent characteristics of OFDM modulation, such as subcarrier orthogonality and robustness to multipath fading, and integrates a secure coding mechanism to design anti-intercept waveforms tailored for satellite networks.
Considering the practical limitations on carrier transmission power in satellite systems, we formulate a secrecy rate maximization problem based on Shannon's secrecy capacity theory, incorporating downlink power constraints to ensure secure and energy-efficient downlink transmission. Additionally, an anti-intercept condition is established, which specifies the minimum difference in signal-to-interference and noise ratio (SINR) required between the legitimate receiver and the eavesdropper, to guide the optimization of the power allocation strategy. This ensures high secrecy rate performance while effectively preventing eavesdropping. The main contributions of this work are as follows:
\begin{itemize}
\item In the downlink transmission scenario of space-ground integrated networks using OFDM modulation, we propose a secrecy rate maximization problem with anti-intercept constraint to address the challenges posed by potential terrestrial Eves. The objective is to strengthen the reception quality for legitimate users while imposing limits on the subcarrier transmission power to achieve secure communication.
\item To solve the non-convex secrecy rate maximization problem with anti-intercept constraint and power constraint, we introduce BSA-Net for secure coding generation and power allocation. First, the original problem is decomposed and simplified into a signal quality maximization problem for legitimate user with power constraint, and we propose an unsupervised learning method for secure coding scheme optimization. Then, a binary search method is proposed to optimize the power allocation scheme to satisfy the anti-intercept constraint. Convergence is achieved through iterative alternation between the two processes.
\item Extensive simulations are conducted to demonstrate the performance and reliability of the proposed approach. Comparative analyses are performed to examine how different factors, such as subcarrier power constraints, subcarrier count, and SINR thresholds at the eavesdropper, affect the downlink secrecy rate. The results demonstrate that the proposed method designs a secure transmission waveform through secure coding scheme, enabling secure and high-speed transmission for legitimate users while preventing Eves from intercepting the communication.
\end{itemize}

The structure of this paper is arranged as follows: Section II reviews existing literature, highlighting both the difficulties and research potential associated with achieving secure and reliable transmission in OFDM-based systems. Section III details the system architecture and formulates the secrecy rate maximization problem, incorporating both power and anti-eavesdropping constraints. In Section IV, we propose BSA-Net, a framework that jointly optimizes secure coding and downlink power allocation scheme to address the proposed problem. Section V showcases the simulation outcomes and provides an in-depth analysis of performance metrics. Finally, Section VI offers concluding remarks and suggests avenues for future research.

\section{Related Work}

In this section, we discuss research and advancements related to secure satellite-terrestrial communication under the background of OFDM modulation technology. Satellite networks have become an pivotal part of the global connectivity envisioned for 6G, serving as an effective supplement for underserved or remote areas \cite{b16}-\cite{b18b}. Prior works such as \cite{b19} and \cite{b20} have investigated multiple dimensions of satellite-based Internet of Things (IoT) systems, focusing on aspects like network architecture, protocol frameworks, and resource allocation. Ensuring the confidentiality of data during transmission has emerged as a critical research priority. Recent studies have highlighted the importance of securing satellite links through encryption schemes, key distribution mechanisms, and reliable routing strategies to preserve data confidentiality and integrity \cite{b20}, \cite{b21}. However, physical layer security, particularly for downlink communication, remains a relatively unexplored domain. 

In recent years, alternative modulation schemes such as orthogonal time frequency space (OTFS) modulation have been proposed. OTFS is considered effective in addressing Doppler shift issues caused by high mobility \cite{b22}. However, it also faces challenges such as high complexity, elevated costs, and difficulties in commercialization. Therefore, OFDM remains the mainstream modulation scheme. As a widely adopted modulation technique, OFDM has been integrated into numerous standards. Hence, studying physical layer security under the OFDM modulation is essential. 

Physical layer security strategies are typically categorized into two distinct paradigms: key-based secrecy methods \cite{b27}, \cite{b28}, and keyless approaches derived from Wyner’s wiretap channel \cite{b29}. Key-based secrecy methods extract pseudo-random sequences from the channel, relying on its reciprocity to ensure that only the intended users can know the sequence. These sequences are used for encrypting data, dynamically interleaving coordinates, or rotating constellations \cite{b30}. Reference \cite{b31} introduces a physical layer encryption method tailored to LoRa modulation, which integrates secret key extraction using received signal strength indicators with a Chinese Remainder Theorem-based secret sharing mechanism. In \cite{b32}, an enhanced physical layer authentication method was introduced, which exploits the phase charactristics of wireless channels and their reciprocal properties for secure information exchange. The secret key information is embedded within the phase of the transmitted signals, and the authentication is achieved through a binary hypothesis testing. \cite{b33} generates dynamic keys by combining traditional keys with the channel characteristics via hashing, leveraging the channel's reciprocity and dynamic nature for encryption. Based on time reversal duplex (TDD) and OFDM systems. \cite{b34} proposed a novel blockchain based key management scheme for Mobile edge computing (MEC), which can effectively resist malicious attacks but still poses the risk of information leakage. In \cite{b35}, a physical layer encryption scheme was designed to alter OFDM data modulation. By transmitting virtual data across multiple OFDM subcarriers and obfuscating data at the subcarrier level, secure transmission with randomized data was achieved. However, key-based physical layer security techniques still require key distribution and authentication, which carries the risk of leakage.

Keyless approaches derived from Wyner's wiretap channel often involves techniques such as adaptive transmission \cite{b36}, \cite{b37}, artificial noise \cite{b38}, and reconfigurable intelligent surfaces (RIS) \cite{b39}, \cite{b40}. \cite{b41} leveraged channel state information (CSI) alongside AI-based methods to optimize user access selection and power allocation, maximizing both satellite-terrestrial link security and reachability rates. A two-stage optimization method is proposed in \cite{b42} to iteratively optimize the beamforming matrix, power allocation scheme, and UAV positioning to maximize the user's secrecy rate performance while ensuring fairness among users. Reference \cite{b43} presents a covert communication architecture utilizing a dual-UAV configuration, in which one UAV operates as a cooperative jammer by generating artificial noise to conceal the presence of covert transmissions and effectively mitigate the risk of detection. In \cite{b44}, RIS technology was introduced into integrated space-air-ground networks, leveraging the channel characteristics of RIS to enhance capability, counter Eves, maximize resource utilization, and ensure communication security. In \cite{b45}, the impact of UAV mobility on secure transmission within ultra-reliable and low-latency communication (URLLC) scenarios was examined. Through strategic adjustment of UAV positioning and critical system parameters, the study ensures user privacy and low-latency requirements while maximizing the system's secrecy rate performance. Keyless physical layer security technology involves complex derivations and computations when addressed using traditional optimization algorithms. Existing research indicates that neural network algorithms exhibit low-complexity properties in implementing keyless physical layer security. However, most of the current work is based on multi-user scenarios, where security transmission is achieved by utilizing noise between users. There is limited research integrating OFDM modulation characteristics with physical layer security techniques, which further inspired our work.

As a keyless secure transmission technique, physical layer security has primarily been investigated from a theoretical perspective in existing research. The majority of existing research emphasizes the enhancement of theoretical indicators like secrecy capacity and secure rate via advanced signal processing strategies, whereas limited attention has been given to achieving a trade-off between transmission security and reliability. Moreover, there has been limited research on achieving secure transmission via physical-layer waveform design.

$Notations:$ \(\otimes\) denotes the cyclic convolution operator. \(\odot\) represents the hadamard product. $\left\|\cdot\right\|$ refers to the Euclidean norm of a vector. The space \(\mathbb{C}^{N\times{N}}\) denotes the set of all \(N\times{N}\) complex-valued matrices. $\mathcal{N}(\mu,\sigma^2)$ represents the Gaussion distribution with mean \(\mu\) and variance \(\sigma^2\). The operator $\left[x\right]^+$ is defined as\(=max(x,0)\), extracting only non-negative values. \(Lin(\cdot)\) represents a real-valued fully connected layer that performs an affine transformation. Additional symbols and variables are summarized in Table~\ref{tab:notations}.

\begin{table}[h!t]
\center
\renewcommand{\arraystretch}{1.5}  
\caption{\small \\SUMMARY OF MAIN NOTATIONS AND DEFINITIONS}
\label{tab:notations}
\begin{tabular}{p{45pt}<{\raggedright}p{185pt}<{\raggedright}}
\hline\\[-3.9mm]\hline
Notation & Definition \\
\hline
\textit{N} & Number of subcarriers\\
\textit{S} & Source Symbol\\
\textit{\({h}\in\mathbb{C}^{N\times{1}}\)} & The legitimate channel from Alice to Bob\\
\textit{\({g}\in\mathbb{C}^{N\times{1}}\)} & The wiretap channel from Alice to Eve\\
\textit{\({m}\in\mathbb{C}^{N\times{N}}\)} & The secure coding matrix\\
\textit{\({p_k}\)} & The power allocated to the $k^{th}$ subcarrier\\
\textit{\(n_b\)} & Noise of the legitimate channel\\
\textit{\(n_e\)} & Noise of the wiretap channel\\
\textit{\({\mathbf{H}}\in\mathbb{C}^{N\times{N}}\)} & The legitimate channel from Alice to Bob in frequency-domain\\
\textit{\({\mathbf{G}}\in\mathbb{C}^{N\times{N}}\)} & The wiretap channel from Alice to Bob in frequency-domain\\
\textit{\({\mathbf{M}}\in\mathbb{C}^{N\times{N}}\)} & The secure coding matrix in frequency-domain\\
\textit{\({\mathbf{P}}\in\mathbb{C}^{N\times{1}}\)} & The column vector represents the transmit power allocated to each subcarrier\\
\textit{\(\delta_k^b\)} & Noise power of the \(k^{th}\) legitimate sub-channel\\
\textit{\(\delta_k^e\)} & Noise power of the \(k^{th}\) wiretap sub-channel\\
\textit{\({\gamma_k^b}\)} & Bob's received SINR on the \textit{\(k^{th}\)} subcarrier\\
\textit{\({\gamma_k^e}\)} & Eve's received SINR on the \textit{\(k^{th}\)} subcarrier\\
\textit{\({R_k}\)} & The secrecy rate of the $k^{th}$ subcarrier\\ 
\textit{\({P_S}\)} & The maximum transmission power for each subcarrier\\
\textit{\({P_b}\)} & The received SER of Bob\\
\textit{\({P_e}\)} & The received SER of Eve\\
\textit{\(\varepsilon_{e}\)} & Constraint of SER of Eve\\
\textit{\(\delta\)} & Tolerance margin\\
\hline
\end{tabular}
\end{table}

\section{System Model And Problem Formulation}


\subsection{System Model}
\begin{figure}[htbp]
\centerline{\includegraphics[width=1.0\linewidth]{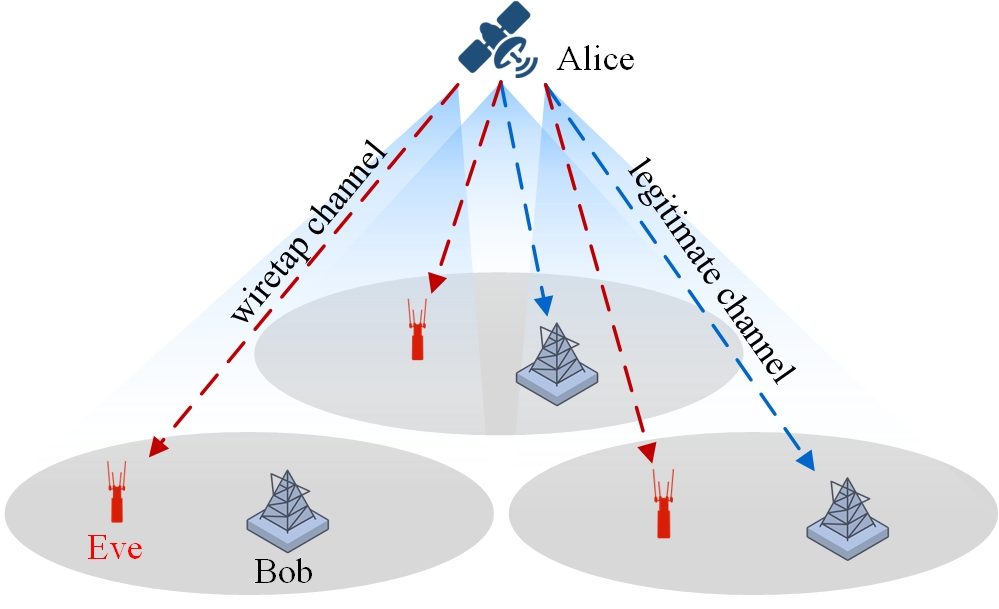}}
\captionsetup{font=small}
\caption{Downlink eavesdropping scenario in satellite networks}
\label{fig-1}
\end{figure}
We investigates secure downlink transmissions in satellite networks shown in Fig. \ref{fig-1}. Assuming a worst-case setting, the channels between the satellite and both the legitimate user and the eavesdropper exhibit similar propagation characteristics. From an information-theoretic security perspective, it is difficult to exploit the inherent stochastic channel variations to ensure secure transmission. Focusing on the widely used OFDM system, the orthogonality among subcarriers results in high correlation between satellite-to-ground subchannels, making secure information transmission still challenging to achieve. Considering the aforementioned challenges, this work focuses on anti-interception secure waveform design for satellite-terrestrial OFDM communication systems.

\begin{figure}[htbp]
\centerline{\includegraphics[width=0.95\linewidth]{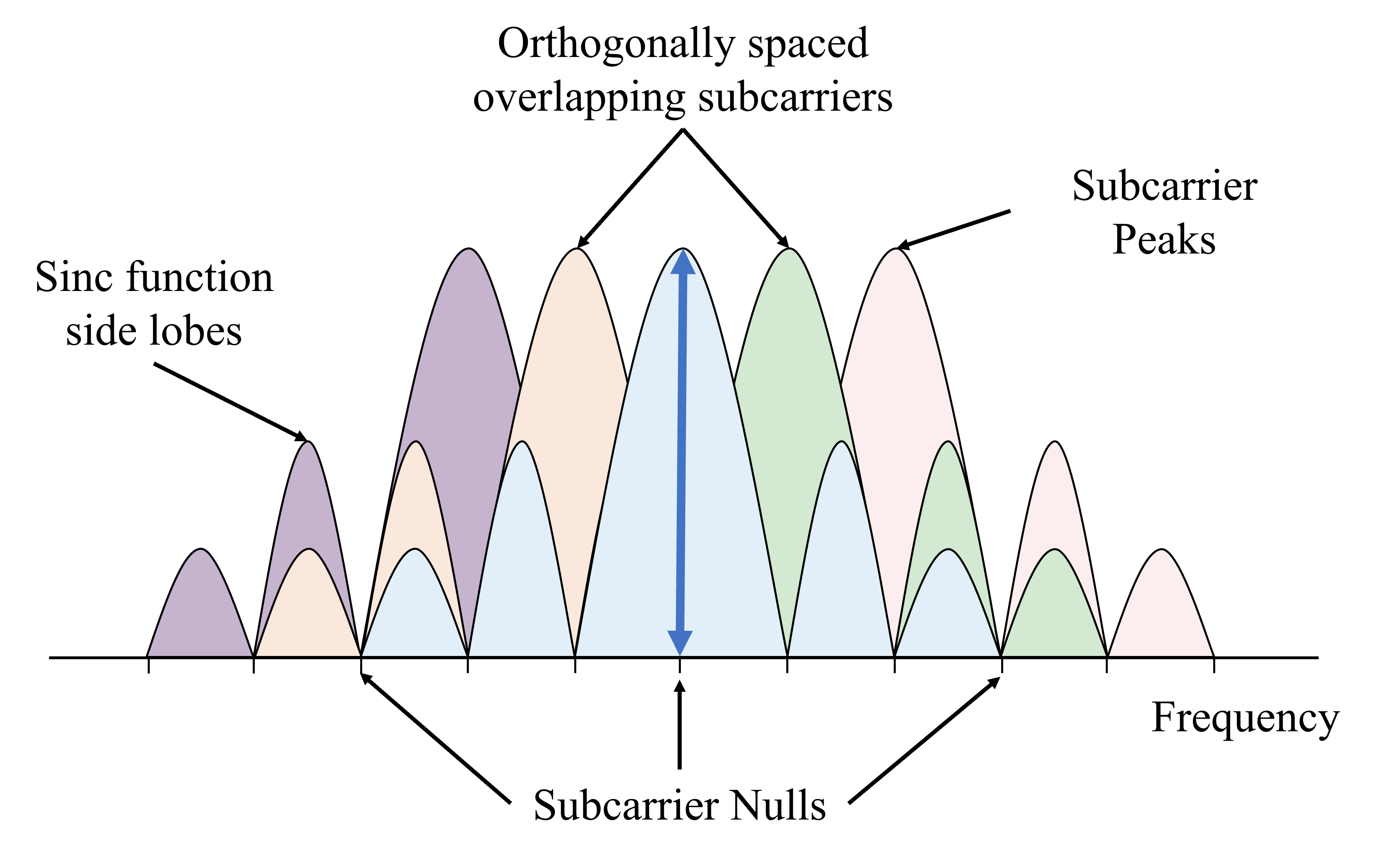}}
\captionsetup{font=small}
\caption{OFDM Signal Frequency Spectrum}
\label{fig-3}
\end{figure}

OFDM system employs Fast Fourier Transform (FFT) and Inverse Fast Fourier Transform (IFFT) operations. The transmitter processes input bits through constellation mapping (e.g., QPSK) to generate complex frequency-domain symbols. These symbols are then fed into an \textit{N}-point IFFT block, where they are converted to time-domain signals through the superposition of \textit{N} orthogonal subcarriers. The complex value of each symbol determines the amplitude and phase information of its respective subcarrier. The resulting time-domain signal represents the summation of all modulated subcarriers. As demonstrated in Fig. \ref{fig-3}, the inherent orthogonality ensures zero inter-carrier interference, with each subcarrier's null points perfectly aligning with other subcarriers' peak frequencies.

We assume the confidential source symbols transmitted across OFDM subcarriers can be represented as
\begin{equation}
     \mathbf{S}=[S_1,S_2,\ldots,S_N]^{T}, \label{eq1}
\end{equation}
and \(\left|S_k\right|^2 = 1\) represents the power of the symbol in the $k^{th}$ subchannel.
Generally, the receiver obtains the signal transmitted through the wireless channel and employs FFT to transform it into the frequency domain for data recovery.

The satellite-to-ground communication channel is modeled as
\begin{equation}
    h = \sqrt{C_L b \beta} \, \exp(-j\theta),
\end{equation}
where $C_L$ denotes the free-space path loss, given by $C_L = (\lambda/4\pi)^2 / (d^2 + l^2)$ with $\lambda$ being the signal wavelength, $d$ the horizontal distance between Alice and Bob, and $l$ the satellite altitude. The term $\beta$ represents rain-induced fading, following a log-normal distribution $\ln(\beta_{\text{dB}}) \sim \mathcal{N}(\mu, \delta^2)$ where $\beta_{\text{dB}} = 10\log_{10}\beta$. The phase shift $\theta$ is uniformly distributed in $[0, 2\pi)$. 

The beam gain $b$ is defined as
\begin{equation}
    b = G_0 \left( \frac{J_1(u_0)}{2u_0} - 36 \frac{J_3(u_0)}{u_0^2} \right)^2,
\end{equation}
where $G_0$ is the maximum antenna gain, $u_0 = 2.07123 \sin(\alpha)/\sin(\alpha_{3\text{dB}})$, $\alpha$ denotes the elevation angle between the beam center and the user, $\alpha_{3\text{dB}}$ is the 3dB beamwidth, and $J_1(\cdot)$, $J_3(\cdot)$ are the first- and third-order Bessel functions of the first kind.


Throughout this paper, lowercase symbols (e.g., $h$, $g$) denote time-domain CSI, while their uppercase counterparts (e.g., $H$, $G$) represent frequency-domain equivalents.


To mitigate eavesdropping threats in satellite-to-ground OFDM communication systems, we propose a waveform design based on secure coding mechanism. Motivated by the channel similarity between legitimate and eavesdropping links, our approach incorporates security encoding prior to wireless transmission through deliberate inter-subcarrier interference generation. Specifically, we design a pre-transmission signal processing technique that creates controlled waveform aliasing across subchannels to obfuscate transmitted symbols against potential Eves. Then the transmitting signal is modified as
\begin{equation}
    \mathbf{x}={\mathbf{p}}\odot\mathbf{m}\otimes{\mathbf{F}}^{-1}{\mathbf{S}},\label{eq4}
\end{equation}
where \({\mathbf{m}}\in\mathbb{C}^{N\times{N}}\) represents the secure coding matrix, \(\otimes\) denotes the cyclic convolution operator, \(\mathbf{S}\in\mathbb{C}^{N\times1}\) is defined by \eqref{eq1}, \({\mathbf{F}}\) is the normalized FFT matrix, and $\mathbf{p}=[\sqrt{p_1},\sqrt{p_2},\ldots,\sqrt{p_N}]^{T}$, where $\sqrt{p_k}$ represents the square root of the transmission power allocated to the $k^{th}$ subcarrier.

Thus, the received signal at Bob and Eve can be respectively written as 
\begin{equation}
    \mathbf{y}_{b}= \mathbf{h}\otimes \mathbf{x}+\mathbf{n}_{b},\label{eq5}
\end{equation}
and
\begin{equation}
    \mathbf{y}_{e}=\mathbf{g}\otimes \mathbf{x}+\mathbf{n}_{e},\label{eq6}
\end{equation}
where $\mathbf{h}\in\mathbb{C}^{N\times 1}$ and $\mathbf{g}\in\mathbb{C}^{N\times 1}$ represent the channel vectors from Alice to Bob and Eve respectively, while $n_b\sim\mathcal{CN}(0,\sigma_b^2)$ and $n_e\sim\mathcal{CN}(0,\sigma_e^2)$ denote the complex additive white Gaussian noise at Bob's and Eve's receivers, with $\sigma_b^2$ and $\sigma_e^2$ being their respective noise variances.

After FFT, the received frequency domain signal vector can be obtained as 
\begin{equation}
    \mathbf{Y}_b=\mathbf{HX}+\mathbf{N}_b,\label{eq7}
\end{equation}
and
\begin{equation}
    \mathbf{Y}_e=\mathbf{GX}+\mathbf{N}_e, \label{eq8}
\end{equation}
where $\mathbf{N}_b\in\mathbb{C}^{N\times 1}$ and  $\mathbf{N}_e\in\mathbb{C}^{N\times 1}$ denote frequency-domain noise, $\mathbf{H}\in\mathbb{C}^{N\times N}$ and $\mathbf{G}\in\mathbb{C}^{N\times N}$ are diagonal matrices, representing frequency-domain channel matrices from Alice to Bob and Eve, respectively. 
Based on \eqref{eq1} and \eqref{eq4}, the secured symbol vector can be expressed as
\begin{equation}
         \mathbf{X}=[\sqrt{p_1}\sum_{n=0}^{N-1}M_{1,n}S_n,,\ldots,\sqrt{p_n}\sum_{n=0}^{N-1}M_{N,n}S_n]^{T}.\label{eq9}
\end{equation}

Through our secure coding design, the received signal model for the anti-intercept OFDM waveform will be reconstructed at the receiver side.Thus, the signal from the \(k^{th}\) sub-channel at Bob can be represented as
\begin{equation}
\begin{aligned}
Y_{k}^{b}& =H_k\sqrt{p_k}\sum_{n=0}^{N-1}M_{k,n}S_n+N_k^{b} \\
&=H_k\sqrt{p_k}M_{k,k}S_k+H_k\sqrt{p_k}\sum_{n=0,n\neq k}^{N-1}M_{k,n}S_n+N_k^{b},\label{eq10}
\end{aligned}
\end{equation}
where $H_k$ denotes the $k^{th}$ diagonal element of the channel matrix $\mathbf{H}$, corresponding to the channel gain of the $k^{th}$ subcarrier, and $N_k^b$ represents the frequency-domain noise on the $k^{th}$ subcarrier received at Bob. 

The Eve wiretaps signal from the \(k^{th}\) subchannel is expressed as
\begin{equation}
\begin{aligned}
Y_{k}^{e}& =G_k\sqrt{p_k}\sum_{n=0}^{N-1}M_{k,n}S_n+N_k^{e} \\
&=G_k\sqrt{p_k}M_{k,k}S_k+G_k\sqrt{p_k}\sum_{n=0,n\neq k}^{N-1}M_{k,n}S_n+N_k^{e}.\label{eq11}
\end{aligned}
\end{equation}
where $G_k$ denotes the $k^{th}$ diagonal element of the channel matrix $\mathbf{G}$, corresponding to the channel gain of the $k^{th}$ subcarrier, and $N_k^e$ represents the frequency-domain noise on the $k^{th}$  subcarrier received at Eve. 

According to \eqref{eq10} and \eqref{eq11}, the SINR at Bob and Eve can be respectively derived as follows:
\begin{equation}
\begin{aligned}
\gamma_k^b&=\frac{p_k\left|H_kM_{k,k}S_k\right|^2}{p_k\left|H_k\sum_{n=0,n\neq k}^{N-1}M_{k,n}S_n\right|^2+{\delta_k^{b}}^2},\label{eq12}
\end{aligned}
\end{equation}
and
\begin{equation}
\begin{aligned}
\gamma_k^e&=\frac{p_k\left|G_kM_{k,k}S_k\right|^2}{p_k\left|G_k\sum_{n=0,n\neq k}^{N-1}M_{k,n}S_n\right|^2+{\delta_k^{e}}^2}.\label{eq13}
\end{aligned}
\end{equation}
where \({\delta_k^{b}}^2\) and \({\delta_k^{e}}^2\) represent the noise power of the \(k^{th}\) legitimate subchannel and eavesdropping subchannel, respectively.

Through analysis of \eqref{eq12} and \eqref{eq13}, we demonstrate that the proposed physical-layer secure coding creates a controlled asymmetric distortion in signal quality between legitimate and eavesdropping receivers. Specifically, 1):
The frequency-selective nature of OFDM channels is exploited to generate subcarrier-dependent SNR degradation at Eve's receiver
2): Symbol-level energy allocation preserves the decoding threshold for legitimate users while intentionally driving Eve's received symbols into nonlinear distortion regions.
3): While the proposed anti-intercept scheme enhances security, it inevitably introduces additional constraints that may degrade the transmission reliability for legitimate users. This critical reliability-security trade-off is also a primary focus of our work.

\subsection{Problem Formulation}
To achieve secure and reliable anti-intercept transmission, we systematically address this pivotal reliability-security tradeoff through our novel optimization approach. The secure coding scheme is designed to exploit the CSI disparity between legitimate and eavesdropping channels, formulated as \(\mathbf{M} = \psi\left(\mathbf{h}_b,\mathbf{h}_e\right)\), where \(\psi\left(\cdot\right)\) is the feature extraction function. Our solution simultaneously resolves: (1) the generation of secure coding matrix and (2) symbol-level power allocation, thereby effectively balancing interception resistance with transmission reliability."

In this work, we quantify reliability performance through dual SER thresholds: the legitimate receiver (Bob) must maintain SER \(\leq \varepsilon_b\), while the Eve should experience SER \(\geq \varepsilon_e\). For QPSK modulated OFDM systems, the SER can be approximated as
\begin{equation}
    \mathrm{SER}(\bar{\gamma})=\frac{1}{2}\mathrm{erfc}(\sqrt{\mathrm{\bar{\gamma}}}),\label{eq14}
\end{equation}
where \(\bar{\gamma}\) means the average SINR.
Besides, utilizing the formulations in \eqref{eq12} and \eqref{eq13}, the secrecy rate of the transmission from Alice to Bob can be derived as
\begin{equation}
    R_k=\left[\log_2\left(1+\gamma_{k}^{b}\right)-\log_2\left(1+\gamma_{k}^{e}\right)\right]^+.\label{eq15}
\end{equation}


\begin{figure*}[hb]
	\centering
	\vspace*{8pt}
	\hrulefill
	\vspace*{8pt} 
	\begin{eqnarray}
        \begin{aligned}
                    R_k&=\left[\sum_{k}\left(\log_{2}\left(1+\frac{p_k\left|H_kM_{k,k}S_k\right|^2}{p_k\left|H_k\sum_{n=0,n\neq k}^{N-1}M_{k,n}S_n\right|^2+{\delta_k^{b}}^2}\right)-\log_{2}\left(1+\frac{p_k\left|G_kM_{k,k}S_k\right|^2}{p_k\left|G_k\sum_{n=0,n\neq k}^{N-1}M_{k,n}S_n\right|^2+{\delta_k^{e}}^2}\right)\right)\right]^{+}\\
        &=\left[\sum_{k}\left(\log_{2}\left(1+\frac{p_k\left|H_kM_{k,k}\right|^2}{p_k\left|H_k\right|^2\sum_{n=0,n\neq k}^{N-1}\left|M_{k,n}\right|^2+{\delta_k^{b}}^2}\right)-\log_{2}\left(1+\frac{p_k\left|G_kM_{k,k}\right|^2}{p_k\left|G_k\right|^2\sum_{n=0,n\neq k}^{N-1}\left|M_{k,n}\right|^2+{\delta_k^{e}}^2}\right)\right)\right]^{+}\label{eq16}.
        \end{aligned}
	\end{eqnarray}
\end{figure*}

To address the critical reliability-security tradeoff, we formulate a secrecy rate maximization problem by generating secure coding matrices and optimizing power allocation across subcarriers, subject to both legitimate user's and Eve's SER constraints. The optimization framework can be expressed as
\begin{align}
&\mathcal{P}1:\underset{{\mathbf{M},\mathbf{p}}}{\operatorname*{\max}} \sum_k R_k \label{eq17}\\
&\text{s.t.:} \quad P_b \leq \varepsilon_b, \tag{\ref{eq17}{a}} \label{eq17a}\\
&P_e \geq \varepsilon_e, \tag{\ref{eq17}{b}} \label{eq17b}\\
&p_k\left|\sum_{n=0}^{N-1}M_{k,n}\right|^2\leq P_S,\forall k, \tag{\ref{eq17}{c}} \label{eq17c}
\end{align}
where \eqref{eq17a} and \eqref{eq17b} are anti-intercept constraints, i.e., \(P_b = SER\left(\bar{\gamma}_k^b\right)\), \(P_e=SER\left(\bar{\gamma}_k^e\right)\), \eqref{eq17c} represent the power constraint for each subcarrier, and \(P_S\) represents the upper limit of the transmission power for each subcarrier.

\section{BSA-Net Based Secure Coding Generation AND Power Allocation}
To solve \(\mathcal{P}1\), we propose a bisection search-activated neural network (BSA-Net) to jointly optimize secure coding and power allocation, maximizing secrecy performance. 
Since \(\mathcal{P}1\) is non-convex due to its high-order and fractional polynomial constraints and objective, the original problem should first be simplified.
Based on the constraint \eqref{eq17b}, \(\mathcal{P}1\) is reformulated as a maximization problem of the transmission rate at Bob's side. Furthermore, based on \eqref{eq12}, \eqref{eq14}, and \eqref{eq17a}, it can be concluded that Bob's communication rate is inversely proportional to its SER. Therefore, the constraint (19a) can be relaxed. Consequently, \(\mathcal{P}1\) is simplified to:
\begin{align}
&\mathcal{P}2:\underset{\mathbf{M}, \mathbf{p}}{\operatorname*{\max}} \sum_k \log_2\left(1+\gamma_k^b\right), \label{eq18}\\
&\text{s.t.:} P_e \geq \varepsilon_e, \tag{\ref{eq18}{a}} \label{eq18a}\\ &{p_k}\left|\sum_{n=0}^{N-1}M_{k,n}\right|^2\leq {P_S},\forall k. \tag{\ref{eq18}{b}} \label{eq18b}
\end{align}

\begin{figure}[htbp]
\centerline{\includegraphics[width=0.95\linewidth]{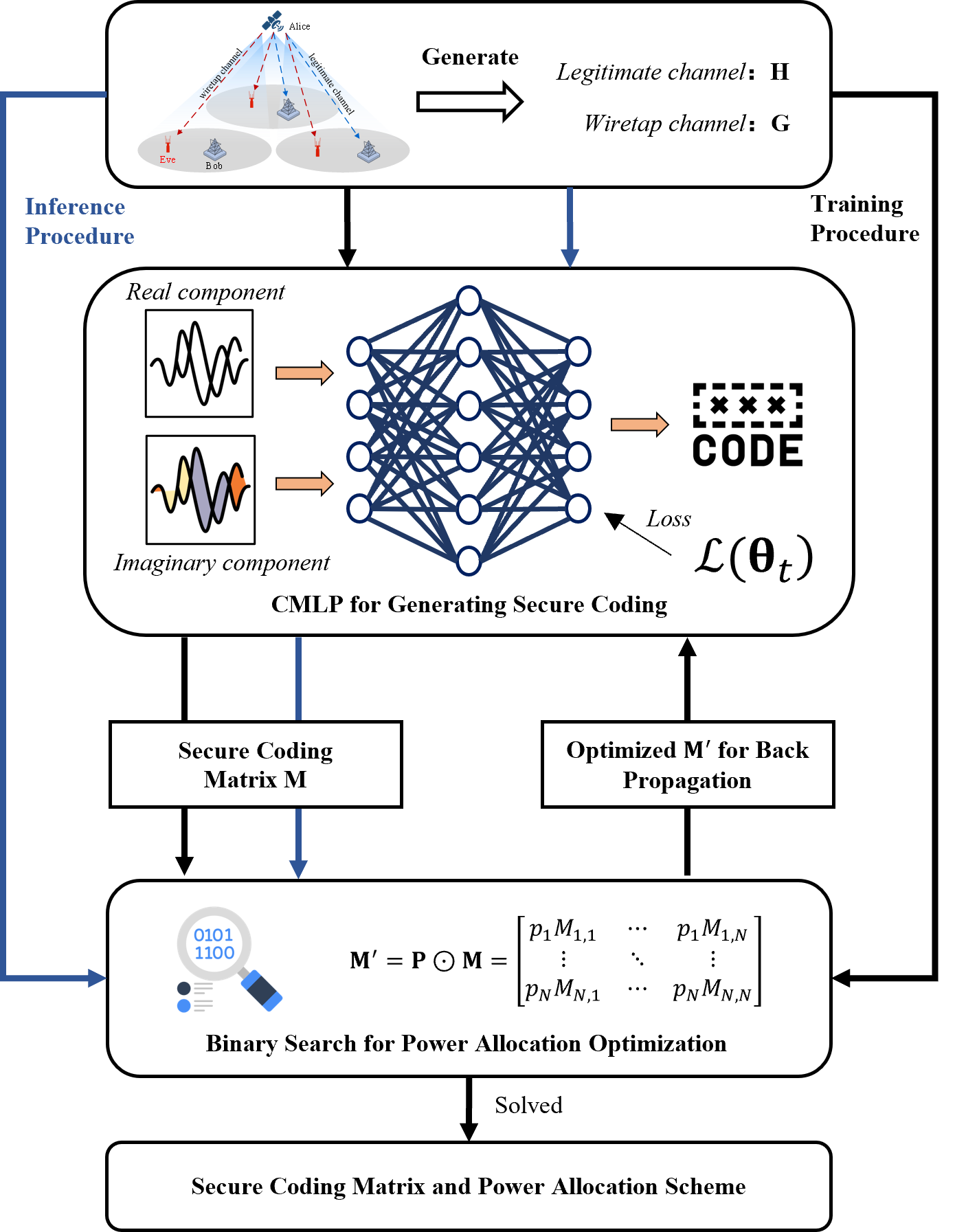}}
\captionsetup{font=small}
\caption{Algorithm Flowchart}
\label{fig-4}
\end{figure}

Next, \(\mathcal{P}2\) is decomposed into two questions for resolution. In the initial phase, the transmission power allocated to subcarriers is fixed to generate the secure coding scheme. Following this, the second phase refines the power allocation scheme among subcarriers, guided by the previously obtained secure coding scheme. Finally, these two optimization phases are alternately executed until the system performance converges. The iterative process of the algorithm is shown in Fig. \ref{fig-4}.

\subsection{Secure Coding Generation}\label{AA}
As indicated by \eqref{eq12}--\eqref{eq14}, and \eqref{eq16}, the SER performance at both Bob and Eve depends on the secure coding scheme and power allocation strategy. Consequently,  we incorporate the power constraint and anti-intercept constraint as respective conditions in the two optimization stages, guaranteeing their simultaneous satisfaction through iterative optimization. To solve the original problem, we initially fix the subcarrier power allocation values \(\mathbf{p}\), thereby reducing the problem to a secure coding optimization problem as follows:
\begin{align}
&\mathcal{P}3:\underset{\mathbf{M}}{\operatorname*{\max}} \sum_k \log_2\left(1+\gamma_k^b\right), \label{eq19}\\
&\left|\sum_{n=0}^{N-1}M_{k,n}\right|^2\leq \frac{P_S}{p_k},\forall k. \tag{\ref{eq19}{a}} \label{eq19a}
\end{align}
\begin{remark}
 This phase implements an unsupervised learning framework for secure coding optimization. To solve the non-convex problem \(\mathcal{P}3\), we design a complex-valued multilayer perceptron (CMLP) that directly processes the native complex representations of both legitimate and wiretap channel parameters. This architecture fundamentally preserves the complete algebraic structure of complex signal spaces, eliminating information degradation inherent in conventional real-valued network paradigms that require ad hoc transformations through extraction or concatenation operations.   
\end{remark}

The neural network accepts as input the composite channel state matrix \(\mathbf{H} = \mathbf{H}_\text{real} + i\mathbf{H}_\text{imag}\), where \(\mathbf{{H}_\text{real}}\) and \(\mathbf{{H}_\text{imag}}\) represent the real and imaginary components, respectively. The CMLP processes is implemented using pairs of real-valued fully connected layers to emulate complex-valued transformations. For any complex input vector, the layer transformation follows the complex-linear mapping:
\begin{equation}
    \mathbf{x}_\text{real}= Lin(\mathbf{H}_\text{real}) - Lin(\mathbf{H}_\text{imag}).  \label{eq20}
\end{equation}
and
\begin{equation}
    \mathbf{x}_\text{imag}=Lin(\mathbf{H}_\text{imag}) + Lin(\mathbf{H}_\text{real}).  \label{eq20b}
\end{equation}
where \(Lin(\cdot)\) represents a real-valued fully connected layer that performs an affine transformation on the input, i.e., \(Lin(\mathbf{x})=\mathbf{Wx}+\mathbf{b}\), where $\mathbf{W}$ and $\mathbf{b}$ are learnable weights and biases, resprectively.

To retain the amplitude and phase correlations inherent in complex-valued data, the CMPL emulate equivalent transformations via traditional real-valued fully connected layers. In our configuration, three complex-valued fully connected layers are utilized, with output dimensions successively set to 128, 64, and 16. Each of these layers incorporates the LeakyReLU activation function to ensure non-linear transformation, which is mathematically expressed as:
\begin{equation}
\text{LeakyReLU}(\mathrm{x}) = \begin{cases}
    \mathrm{x}, & \text{if } \mathrm{x} > 0 \\
    \sigma \mathrm{x}, & \text{if } \mathrm{x} \leq 0 \label{eq21}
\end{cases}
\end{equation}
where \(\sigma\) is a small constant (typically between 0.01 and 0.1), which controls the slope of the activation function for negative input values, which maintains a non-zero gradient for negative inputs, thereby alleviating the gradient vanishing problem typically encountered with standard ReLU.

The training of the network is carried out using the Adam algorithm, which dynamically adapts the learning rate based on first- and second-order moment estimates of the gradient, thereby facilitating robust and efficient convergence. The network outputs secure coding scheme for OFDM systems. The update of neural parameters  \(\theta\) at iteration \(t\) follows the rule: 
\begin{equation}
\theta_t = \theta_{t-1} - \alpha \frac{\hat{m}_t}{\sqrt{\hat{v}_t} + \epsilon}, \label{eq22}
\end{equation}
where \(\epsilon\) is a small constant to avoid division by zero, \(\alpha\) denotes the learning rate, and \(\hat{m}_t\) and \(\hat{v}_t\) are bias-corrected estimates of the first and second moments, respectively, defined as:
\begin{equation}
\hat{m}_t = \frac{m_t}{1 - \beta_1^t}, \quad \hat{v}_t = \frac{v_t}{1 - \beta_2^t}, \label{eq23}
\end{equation}
with
\begin{equation}
m_t = \beta_1 m_{t-1} + (1 - \beta_1) \mathcal{L}(\boldsymbol{\theta}_t), \label{eq24}
\end{equation}
and
\begin{equation}
v_t = \beta_2 v_{t-1} + (1 - \beta_2) \mathcal{L}(\boldsymbol{\theta}_t)^2. \label{eq25}
\end{equation}
where \(\beta_1\) and \(\beta_2\)  are exponential decay rates controlling the momentum of the mean and variance, respectively, and \(\mathcal{L}(\boldsymbol{\theta}_t)\) represents the loss gradient at step \(t\).

Given the power constraint in  the optimization problem \(\mathcal{P}3\) (as expressed in \eqref{eq19a}), a natural approach is to employ Lagrangian dual method, thereby converting the constrained problem into an unconstrained formulation. The neural network (NN) is then trained to optimize the dual Lagrangian objective. However, due to the statistical learning nature of neural networks, which inherently lack strict constraint interpretability, the model can only achieve a probabilistic approximation of the constrained solution. As a result, there is no guarantee that the NN outputs will always satisfy the predefined constraint.

To address this problem, we introduce a specially designed activation function, denoted by \(\mathcal{S}\), to ensure that the NN output adheres to the power constraint. The function \(\mathcal{S}\) operates on complex-valued vectors to ensure the square of the complex vector do not exceed a given power threshold \(P_S\). Specifically, if the squared norm of  the complex vector falls within the limit, the input is passed through unchanged; otherwise, it is scaled proportionally to project it onto the feasible set. The function \(\mathcal{S}\) is formally defined as:
\begin{equation}
\mathcal{S}(\mathbf{M}_k) = \begin{cases}
\mathbf{M}_k, & \text{if } \|\mathbf{M}_k\|^2 \leq P_S \\
\mathbf{M}_k \frac{\sqrt{P_S}}{\|\mathbf{M}_k\|}, & \text{if } \|\mathbf{M}_k\|^2 > P_S \label{eq26}
\end{cases}
\end{equation}
where $\mathbf{M}_k$ represents the $k^{th}$ row of secure coding matrix $\mathbf{M}$, including all column elements. 

To ensure back propagation compatibility, we derive the Jacobian of \(\mathcal{S}\) with respect to the input complex-valued vector \(\mathbf{M}_k\):
\begin{equation}
\frac{\partial \mathcal{S}(\mathbf{M}_k)}{\partial \mathbf{M}_k} = \begin{cases}
1, & \text{if } \|\mathbf{M}_k\|^2 \leq P_S \\
\frac{\sqrt{P_S}}{\|\mathbf{M}_k\|} - \frac{\mathbf{M}_k (\mathbf{M}_k^H \mathbf{M}_k)}{\|\mathbf{M}_k\|^3}, & \text{if } \|\mathbf{M}_k\|^2 > P_S \label{eq27}
\end{cases}
\end{equation}

This piecewise definition ensures differentiability of \(\mathcal{S}\), enabling gradient-based optimization. Incorporating the activation function \(\mathcal{S}\), we define the loss function as:
\begin{equation}
    \mathcal{L}(\boldsymbol{\theta}_t)=-\sum_k \log_2\left(1+\gamma_k^b\right). \label{eq28}
\end{equation}

This formulation allows for back propagation through NN, supporting unsupervised learning while ensuring the power constraint is respected. To further reinforce the magnitude constraint on the secure coding generation, we define a mask as follows:
\begin{equation}
\mathrm{mask}_k = \max\left(1, \frac{\|\mathbf{M}_k\|^2}{P_S}\right). \label{eq29}
\end{equation}
and apply it as:
\begin{equation}
{\mathbf{M}_{k}}^{\prime} = \frac{{\mathbf{M}_{k}}}{\mathrm{mask}_k}. \label{eq30}
\end{equation}
guaranteeing that the power constraint is satisfied for each subcarrier’s encoding vector.

By integrating the activation function \(\mathcal{S}\) and the loss function \(\mathcal{L}\), the proposed NN framework not only maintains compliance with the power constraint but also facilitates effective secure encoding for OFDM systems. The overall structure allows the gradients of the objective function with respect to network parameters to be computed via the chain rule.

It is worth noting that, as indicated by \eqref{eq13}, the SINR at Bob depends on the power allocation scheme. Thus, before applying back propagation in \eqref{eq28}, we must first determine a power allocation strategy that meets the anti-intercept constraint in \eqref{eq18a}. This motivates the following subsection.

\subsection{Power Allocation Optimization}
Building upon the secure coding scheme derived through unsupervised learning, we propose an iterative power allocation optimization method inspired by the binary search algorithm.

Under the constraints imposed by \eqref{eq13}, \eqref{eq14}, and \eqref{eq18a}, the power allocation scheme must be optimized to satisfy both the transmit power budget and the anti-interception requirement. By exploiting the optimized secure coding scheme and the structure of the matrix \(\mathbf{S}\), the per-subcarrier Eve SINR can be expressed as:
\begin{equation}
    \gamma_k^e = \frac{p_k\left|H_k^eM_{k,k}^{\prime}\right|^2}{p_k\left|H_k^e\right|^2\sum_{n=0,n\neq k}^{N-1}\left|M_{k,n}^{\prime}\right|^2+{\delta_k^{e}}^2}, \label{eq31}
\end{equation}
where \(p_k\) represents the transmission power allocated to the \(k^{th}\) subcarrier, and $M_{k,k}^{\prime}$ represents the elment in the $k^{th}$ row and $k^{th}$ column of the optimized secure coding $\mathbf{M^{\prime}}$.

Based on \eqref{eq14} and \eqref{eq31}, the objective of power allocation optimization is to maximize the power allocation scheme while satisfying the anti-intercept constraint:
\begin{align}
&\mathcal{P}4:\max \sum_kp_k,\label{eq32}\\
&\min_{k} \mathrm{SER}\left({\gamma_k^e}\right)\geq\varepsilon_e,\forall k.\tag{\ref{eq32}{a}} \label{eq32a}\\
&p_k\leq \frac{P_S}{\left|\sum_{n=0}^{N-1}M_{k,n}\right|^2},\forall k.\tag{\ref{eq32}{b}} \label{eq32b}
\end{align}

Instead of calculating the SER performance based on the average SINR at Eve, a stricter anti-intercept constraint is adopted in power allocation optimization. This ensures that the SER performance of all subcarriers at Eve exceeds \(\varepsilon_e\), effectively satisfying \eqref{eq18a}. 

Therefore, adjustable power factors $\mathbf{p}=[\sqrt{p_1},\ldots,\sqrt{p_N}]^{T}$ initially set to $\frac{P_S}{\left|\sum_{n=0}^{N-1}M_{k,n}^{\prime}\right|^2}$ for all subcarriers. Based on the wiretap channel parameters and the secure coding scheme, the initial SER performance of each subcarrier at Eve is calculated. To determine whether the SER performance of each subcarrier at Eve satisfies the anti-intercept constraint, a tolerance threshold is set. The condition is met when the maximum deviation of the SER performance of all subcarriers at Eve from the target value falls below the tolerance threshold, i.e.,
\begin{equation}
    \left| \min_{k} \mathrm{SER}_{k}^{e} - \varepsilon_{e} \right| \leq \delta, \label{eq33}
\end{equation}
where \(\mathrm{SER}_{k}^{e}\) represents the SER of the \(k^{th}\) subcarrierat Eve, \(\varepsilon_{e}\) is the target SER value, and \(\delta\) is the predefined tolerance threshold. 

To solve \(\mathcal{P}4\), we calculate the SINR values for all subchannels at Eve and select the subchannel \(k\) that has the highest SINR and whose SER performance violates the constraint in \eqref{eq33}. For the identified subchannel \(k\), employ a binary search method to optimize its power allocation scheme $\mathbf{p}$ within the range (0,$\frac{P_S}{\left|\sum_{n=0}^{N-1}M_{k,n}^{\prime}\right|^2}$]. The binary search process is shown in Algorithm 1.

\begin{algorithm}[!ht]
    \caption{Bisection Search for Adjustable Power Factors with Repeat Loop}
    \label{alg:AOS}
    \renewcommand{\algorithmicrequire}{\textbf{Input:}}
    \renewcommand{\algorithmicensure}{\textbf{Output:}}
    
    \begin{algorithmic}[1]
        \REQUIRE \(\mathbf{H}\), \(\mathbf{G}\), \(\mathbf{M}\), \(\varepsilon_e\), \(\delta\), max\_iterations  
        \ENSURE \(\mathbf{p}\)   
        
        \STATE Initialize \(p_k \gets \frac{P_S}{\left|\sum_{n=0}^{N-1}M_{k,n}^{\prime}\right|^2}, \forall k\) \label{line:init}
        \STATE Calculate SER of \(k^{th}\) subchannel at Eve based on \eqref{eq14}: \(\text{index} = \arg\min_{k}(\mathrm{SER}_k^e)\)
        
        \WHILE{\(\left|\mathrm{SER}_\text{index}^{e} - \varepsilon_{e} \right| \geq \delta\)} \label{line:while_start}
            \STATE Compute \(p_{\text{mid}} \gets (p_{\text{min}} + p_{\text{max}}) / 2\) \label{line:midpoint}
            
            \REPEAT \label{line:repeat_start}
                \IF{\(\mathrm{SER}_\text{mid}^e > \varepsilon_e\)} \label{line:check_ser}
                    \STATE Update \(p_{\text{min}} \gets p_{\text{mid}}\), \(p_{\text{index}} \gets p_{\text{mid}}\)
                \ELSE
                    \STATE Update \(p_{\text{max}} \gets p_{\text{mid}}\), \(\)
                \ENDIF
                
                \STATE Increment iteration counter: \(iter \gets iter + 1\)
                
            \UNTIL{\(\left|\mathrm{SER}_\text{index}^{e} - \varepsilon_e \right| < \delta\) \text{ or } \(iter \geq \text{max\_iterations}\)} \label{line:repeat_end}
            \STATE Update the index corresponding to the minimum SER \(\text{index} = \arg\min_{k}(\mathrm{SER}_k^e)\)
        \ENDWHILE \label{line:while_end}
        
        \RETURN \(\mathbf{p}\)
    \end{algorithmic}
\end{algorithm}

Given the power factors output from Algorithm 1, the loss function in \eqref{eq28} can be further expressed as
\begin{equation}
    \mathcal{L}(\boldsymbol{\theta}_t)=-\sum_k \log_2\bigg(1+\frac{p_k\left|H_k^bM_{k,k}^{\prime}\right|^2}{p_k\left|H_k^b\right|^2\sum_{n=0,n\neq k}^{N-1}\left|M_{k,n}^{\prime}\right|^2+{\delta_k^{b}}^2}\bigg). \label{eq34}
\end{equation}

This loss function, which incorporates the optimized power factors \(p_k\), is employed during backpropagation to update the neural network’s weights and parameters in the secure coding optimization model.

Based on the above work, the BSA-Net algorithm for jointly optimizing secure coding scheme and power allocation is presented in Algorithm 2.
\begin{algorithm}[!ht]
    \caption{BSA-Net for Joint Optimization of Secure Coding Scheme and Power Allocation}
    \label{alg:JointOptimization}
    \renewcommand{\algorithmicrequire}{\textbf{Input:}}
    \renewcommand{\algorithmicensure}{\textbf{Output:}}
    
    \begin{algorithmic}[1]
        \REQUIRE \(\mathbf{H}\), \(\mathbf{G}\), \(\mathbf{M}\), \(\varepsilon_e\), \(\delta\), max\_iterations
        \ENSURE \(\mathbf{M}\), \(\mathbf{p}\)
        \STATE \textbf{Training Procedure:}
        \STATE \textbf{Secure Coding Optimization:}
        \FOR{each epoch}
            \STATE Input \(\mathbf{H}\) and \(\mathbf{G}\) for secure coding optimization;
            \STATE Obtain corresponding secure coding \(\mathbf{M}\)
            \STATE \textbf{Power Allocation Optimization:}
            \STATE \quad Based on \(\mathbf{M}\), \(\mathbf{H}\) and \(\mathbf{G}\), execute \(\textbf{Algorithm 1}\) for power allocation optimization.
            \STATE Calculate the loss function by \eqref{eq34} and update the NN's parameters by \eqref{eq22} using Adam.
        \ENDFOR
        \STATE \textbf{Inference Procedure:}
        \STATE \quad 1) Input \(\mathbf{H}\) and \(\mathbf{G}\) for secure coding optimization;
        \STATE \quad 2) NN for secure coding optimization outputs the optimal secure coding scheme \(\mathbf{M}\) with power constraint;
        \STATE \quad 3) Input \(\mathbf{M}\), \(\mathbf{H}\) and \(\mathbf{G}\) for power allocation optimization. 
        \STATE \quad 4) \(\textbf{Algorithm 1}\) outputs the optimal power allocation acheme \(\mathbf{p}\) with anti-intercept constraint.
        \STATE \quad 5) Secure coding \(\mathbf{M}\) and power allocation scheme \(\mathbf{p}\) jointly achieve waveform design with anti-interception capability.
    \end{algorithmic}
\end{algorithm}

\section{Performance Evaluations}
This section presents simulation experiments designed to assess the secrecy performance of downlink communications. The simulation parameters are configured as follows: the satellite transmitter (Alice) operates at an orbital altitude of 600km, while the ground base station (Bob) is randomly located within an 800km radius centered at the coverage area orgin. The Eve is also located on the ground, with its coordinates randomly distributed within a radius of 1000 km from Alice. At a reference distance of 1 m, the channel power gains from the satellite to the Bob and to the ground Eve are set to 10 dB and 5 dB, respectively. For the downlink scenario, the propagation between satellite and ground receiver is modeled with a Rician factor of 10 dB.





\begin{figure}[htbp]
\centerline{\includegraphics[width=1.0\linewidth]{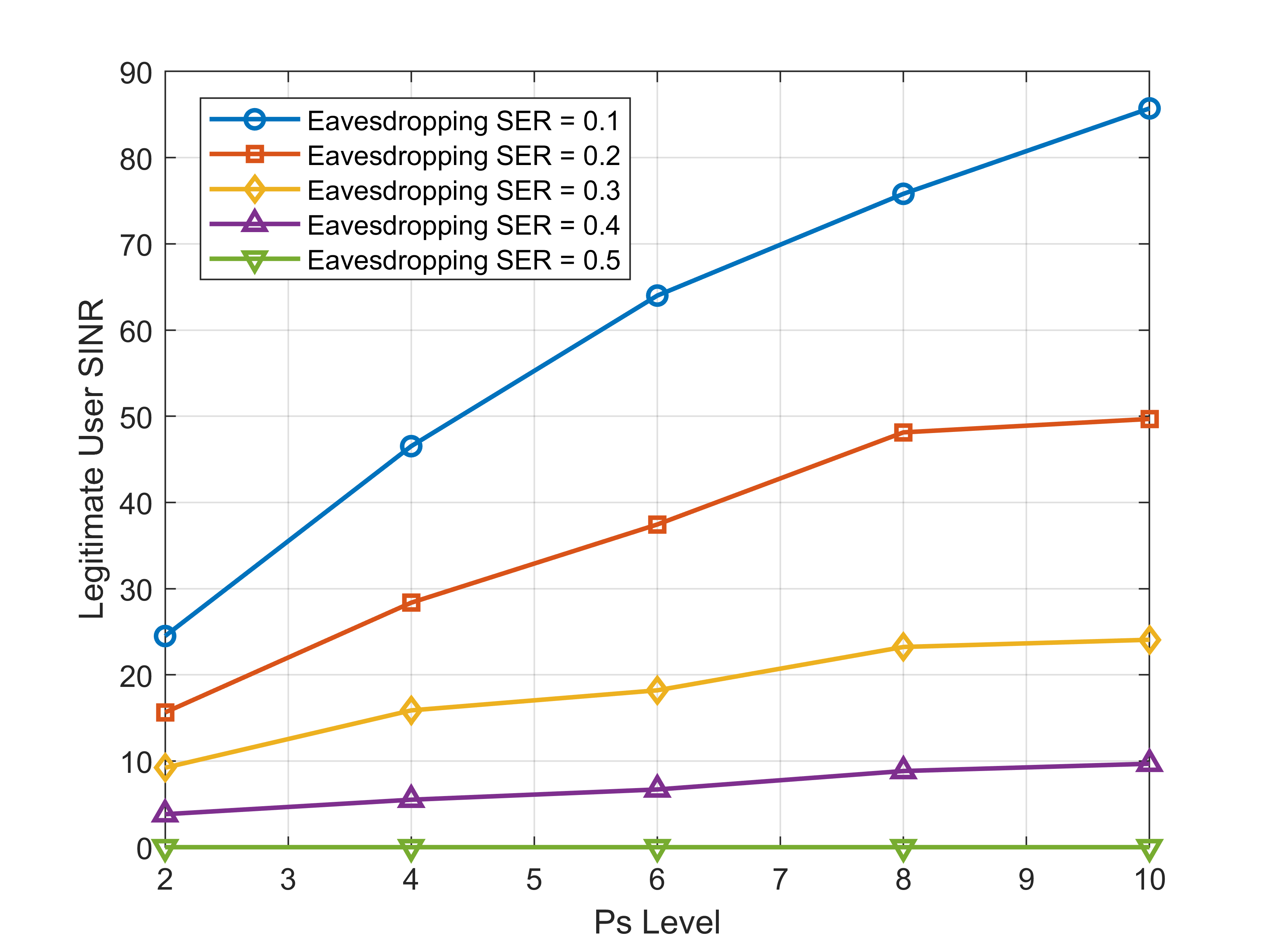}}
\captionsetup{font=small}
\caption{The impact of satellite transmission power on the legitimate user SINR}
\label{fig-5}
\end{figure}

Fig. \ref{fig-5} demonstrates how varying transmit power influence the SINR performance of the legitimate receiver. As the power constraint on each subcarrier increases, the SINR correspondingly improves, owing to the enhanced main channel quality achieved under the imposed anti-interception conditions. Nonetheless, the rate of SINR growth tends to plateau at elevated power levels, which can be attributed to two primary reasons:
Noise-limited regime: At lower power levels, environmental noise dominates, limiting Eve’s symbol error rate (SER). Available power primarily improves the signal quality of legitimate user.
Interference allocation: As power increases, more energy must be allocated to inter-carrier interference (ICI) to degrade Eve’s SER, causing the legitimate SINR gain to saturate.
Equations \eqref{eq12} and \eqref{eq13} reveal a fundamental trade-off: stricter anti-interception constraints (lower Eve’s SINR) reduce the legitimate user’s SINR, as more power is diverted to ICI generation. When Eve’s SER reaches 0.5 (no information extraction possible), the useful information rate in legitimate channel also collapses – a deliberate security-driven degradation.

\begin{figure}[htbp]
\centerline{\includegraphics[width=1.0\linewidth]{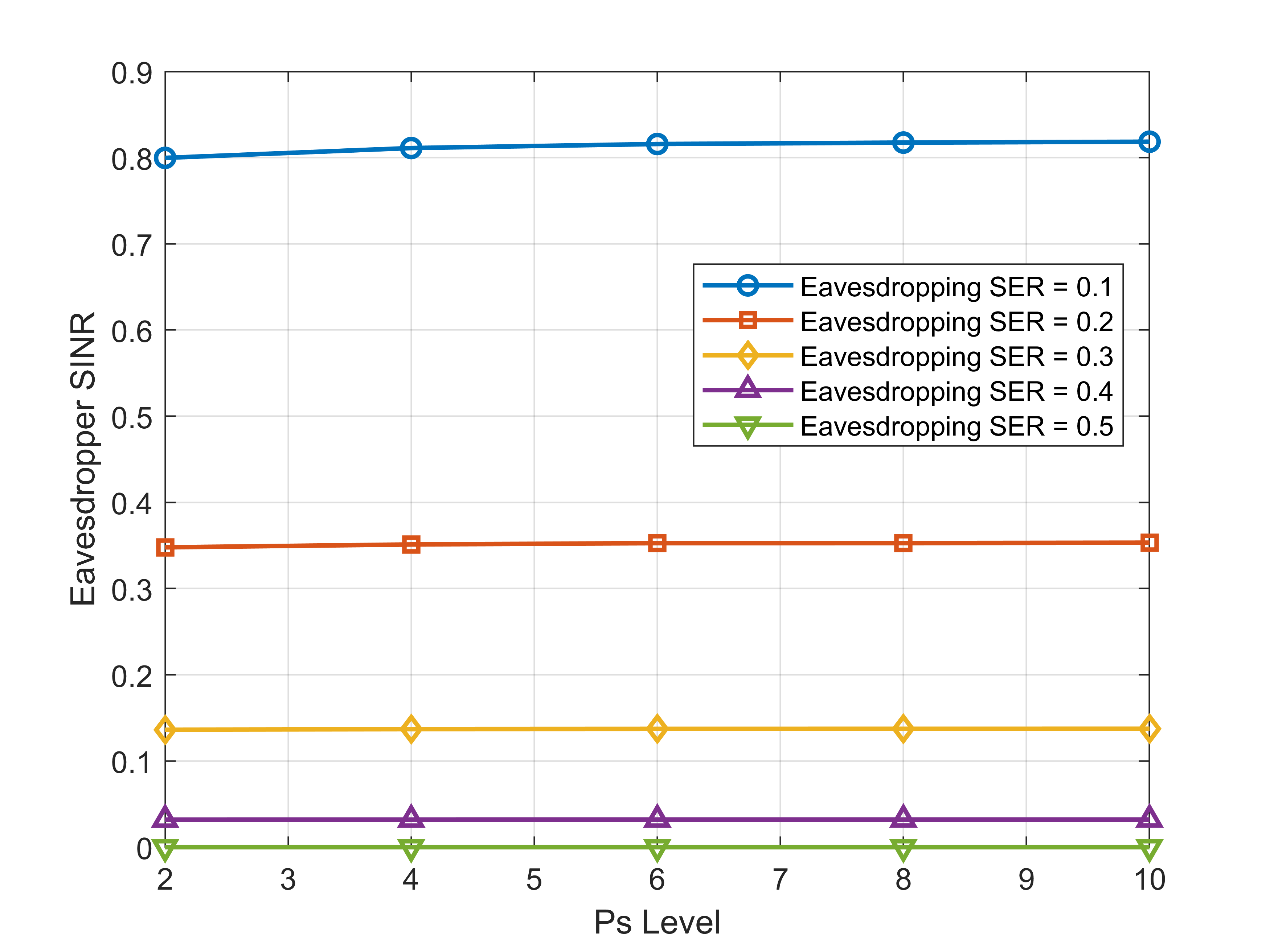}}
\captionsetup{font=small}
\caption{The impact of satellite transmission power on the SINR of Eve }
\label{fig-6}
\end{figure}

Fig. \ref{fig-6} further demonstrates the SINR performance at Eve's end under different anti-intercept constraints. It can be observed that, regardless of the transmission power upper limit, the SINR performance at Eve's side is constrained within a certain range. After our calculations, it is found that the corresponding SER performance closely matches the anti-intercept constraints. This proves that the proposed two-stage optimization method is effective in optimizing the power allocation scheme to limit the Eve's SINR performance and satisfy the anti-intercept constraints.
\begin{figure}[htbp]
\centerline{\includegraphics[width=1.0\linewidth]{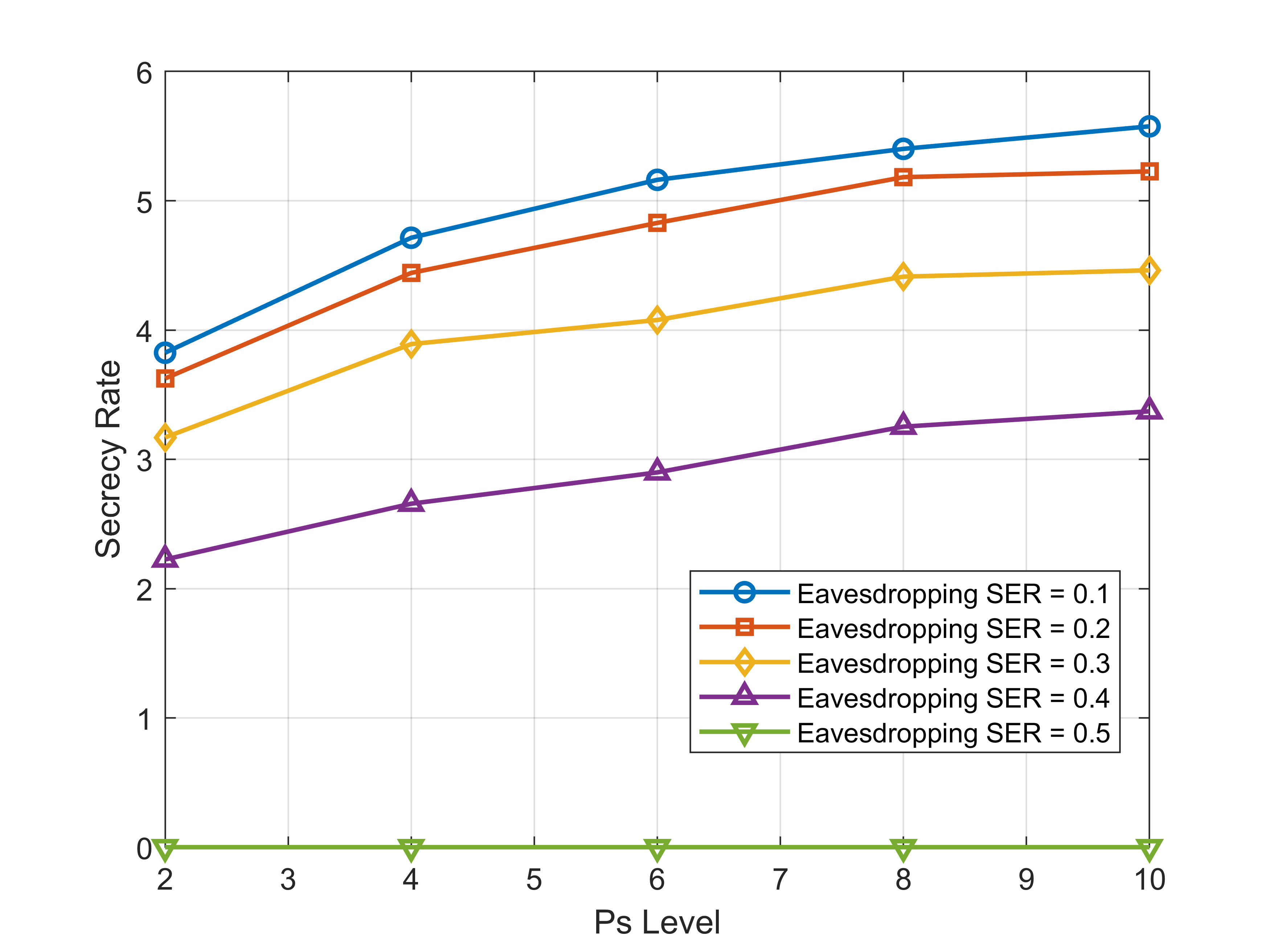}}
\captionsetup{font=small}
\caption{The impact of satellite transmission power on the average secrecy rate}
\label{fig-7}
\end{figure}

Fig. \ref{fig-7} demonstrates the impact of the upper bound of subcarrier's transmission power on the secrecy rate performance. It is evident that as the upper bound of the transmission power increases, the secrecy rate performance shows an increasing trend, which is consistent with the changes in SINR performance at both the legitimate user's end and the Eve's end as the transmission power upper limit increases. When the anti-intercept constraint is set to 0.4, a high secrecy communication rate can be effectively achieved while ensuring anti-intercept performance. However, when the SER is set to 0.5, although the Eve is fully interfered with, the legitimate user is also unable to obtain any useful information, resulting in a secrecy rate performance of 0.
\begin{figure}[htbp]
\centerline{\includegraphics[width=1.0\linewidth]{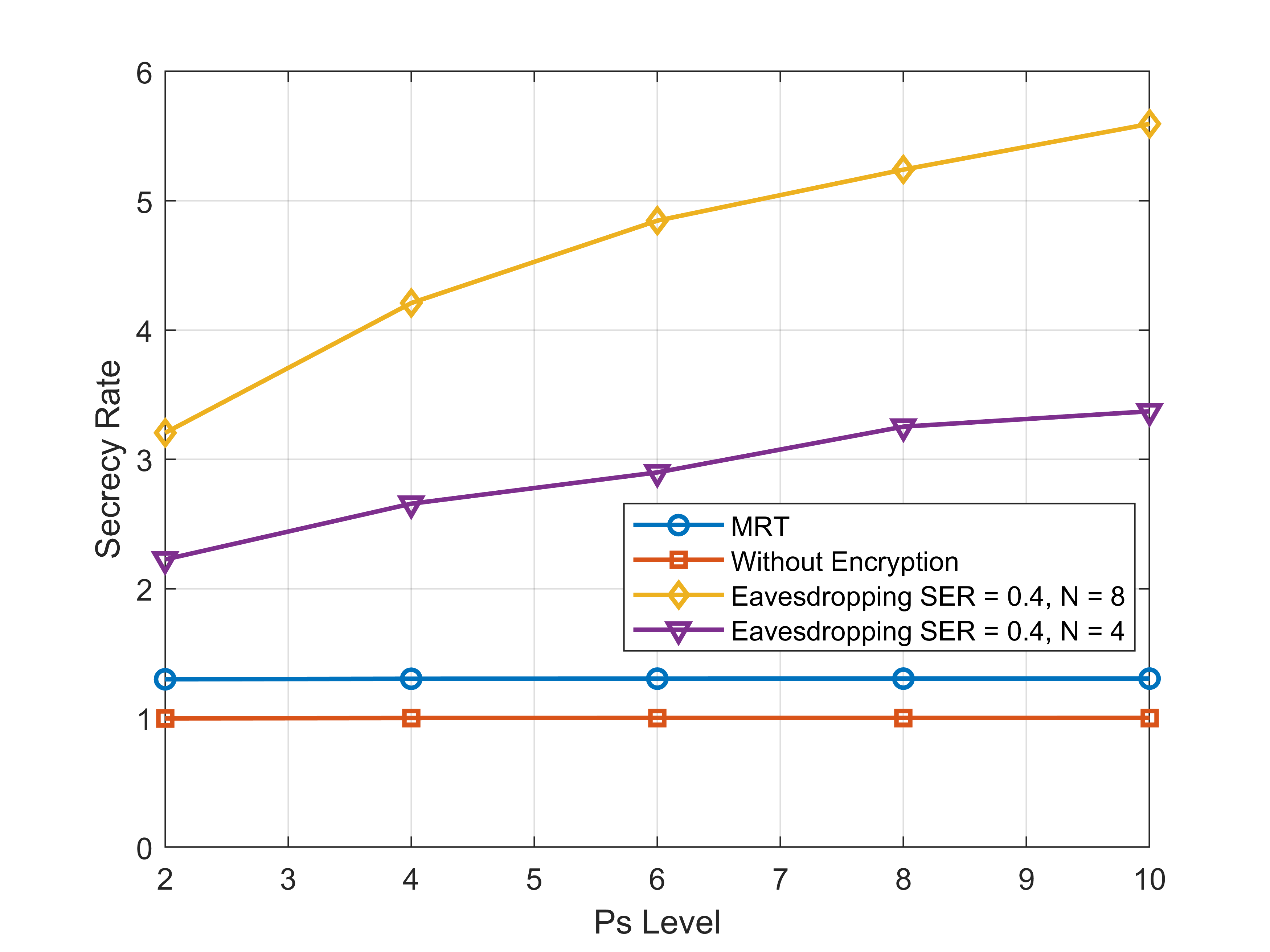}}
\captionsetup{font=small}
\caption{Comparative Trials on the impact of satellite transmission power on the average secrecy rate}
\label{fig-8}
\end{figure}

Fig. \ref{fig-8} illustrates how the secrecy rate varies with the maximum transmit power constraint under different transmission schemes. As shown, the secrecy rate performance is significantly lower in both the unencrypted and MRT (Maximum Ratio Transmission) methods. Their secrecy rate performance is relatively poor because the scheme's performance is primarily dependent on the CSI. Even if the upper limit of the transmit power increases, both the legitimate user's and the Eve's SINR performance improve, resulting in little impact on the secrecy rate performance. In contrast, the proposed method substantially enhances the secrecy rate performance. Additionally, it can be observed that a higher number of subcarriers results in a better secrecy rate performance. This trend can be explained analytically: a greater number of subcarriers increases the denominator terms in the formula, which means that for achieving the same anti-eavesdropping effect, less interference is required among subcarriers when there are more subcarriers, as the denominator comprises the sum of inter-subcarrier interference and environmental noise.
\begin{figure}[htbp]
\centerline{\includegraphics[width=1.0\linewidth]{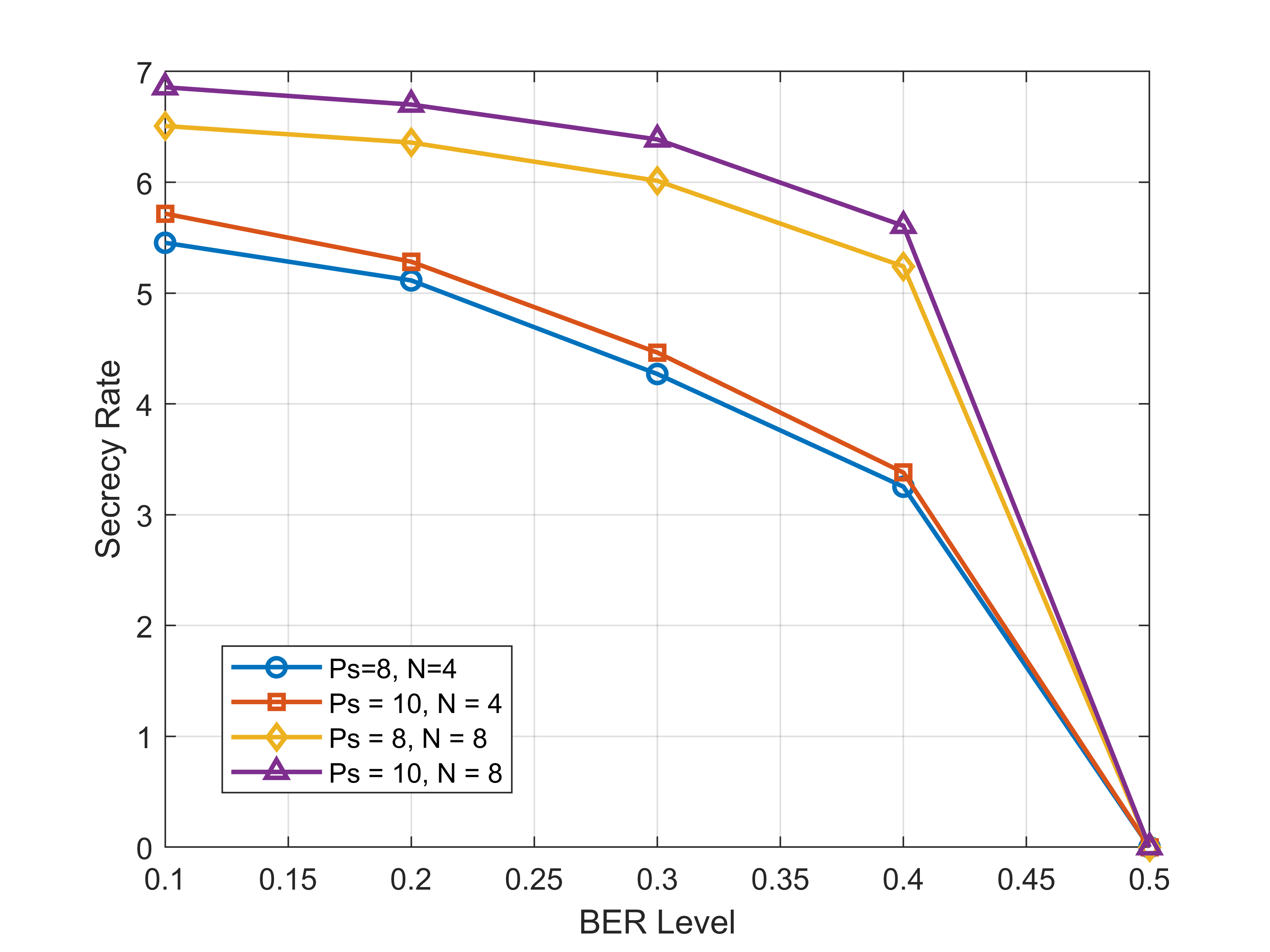}}
\captionsetup{font=small}
\caption{The impact of SER constraint on the secrecy performance}
\label{fig-9}
\end{figure}

Fig. \ref{fig-9} further demonstrates the impact of the anti-intercept constraints on secrecy rate performance. As can be observed, the secrecy rate performance decreases as the anti-intercept constraint becomes more stringent. This is because a higher anti-intercept constraint requires more power to be allocated as interference in order to prevent eavesdropping, which, under the same transmission power limit, reduces the SINR performance at the legitimate user's end. Consequently, the secrecy rate performance also declines. On the other hand, increasing the number of subcarriers results in better secrecy rate performance. This is because with more subcarriers, the power required for interference on each individual subcarrier is relatively smaller. As a result, the SINR performance for each subcarrier at the legitimate user’s end improves, which in turn leads to an increase in the overall secrecy rate performance. By integrating the characteristics of OFDM modulation with the principles of physical layer security, our method strikes a balance between power efficiency and security requirements.

\section{Conclusion}
This paper addresses the challenges of secure downlink transmission in space-ground integrated networks by proposing a novel physical layer secure coding scheme tailored for OFDM systems. By leveraging the characteristics of OFDM modulation, we designed a secure transmission waveform through coding mechanism and  power allocation optimization. The proposed BSA-Net not only maximizes the secrecy rate under transmission power constraints but also incorporates anti-intercept constraints to effectively enhance resistance against malicious attacks. Additionally, the method is highly flexible, capable of adapting to dynamic satellite network conditions, making it a promising solution for secure communication in next-generation 6G satellite-integrated networks. Future research will further extend this framework to multi-satellite and multi-user scenarios to enhance its applicability in complex and dynamic communication environments.
\ifCLASSOPTIONcaptionsoff
  \newpage
\fi


\begin{thebibliography}{00}
\bibitem{b1} N. Cheng, W. Quan, W. Shi, H. Wu, Q. Ye, H. Zhou, W. Zhuang, X. Shen, and B. Bai, "A Comprehensive Simulation Platform for Space-Air-Ground Integrated Network," \emph{IEEE Wireless Commun.}, vol. 27, no. 1, pp. 178-185, Feb. 2020.
\bibitem{b2b} D. Li, X. Liu, Z. Yin, N. Cheng, and J. Liu, "CWGAN-Based Channel Modeling of Convolutional Autoencoder-Aided SCMA for Satellite-Terrestrial Communication," \emph{IEEE Internet of Things Journal}, vol. 11, no. 22, pp. 36775-36785, 15 Nov.15, 2024.
\bibitem{b2} Propagation Model for IF77, P Series, ITU-R Rep., Geneva, Switzerland, 2020, pp. 1–72. [Online]. Available: http://www.itu.int/publ/RREP/en
\bibitem{b3} R. Deng, B. Di, H. Zhang, L. Kuang and L. Song, "Ultra-Dense LEO Satellite Constellations: How Many LEO Satellites Do We Need?," \emph{IEEE Trans. Wireless Commun.}, vol. 20, no. 8, pp. 4843-4857, Aug. 2021.
\bibitem{b4} H. Lim, J. Lee, J. Lee, S. D. Sathyanarayana, J. Kim, A. Nguyen, K. T. Kim, Y. Im, M. Chiang, D. Grunwald, K. Lee, and S. Ha, "An Empirical Study of 5G: Effect of Edge on Transport Protocol and Application Performance," \emph{IEEE Trans. Mob. Comput.}, vol. 23, no. 4, pp. 3172-3186, Apr. 2024.
\bibitem{b5} Y. Zhang, A. Doshi, R. Liston, W. Tan, X. Zhu, J. Andrews, and R. Heath "DeepWiPHY: Deep Learning-Based Receiver Design and Dataset for IEEE 802.11ax Systems," \emph{IEEE Trans. on Wireless Commun.}, vol. 20, no. 3, pp. 1596-1611, Mar. 2021.
\bibitem{b7} J. Cheng, S. Guo and J. He, "An Extended Type-1 Generalized Feistel Networks: Lightweight Block Cipher for IoT," \emph{IEEE Internet Things J.}, vol. 9, no. 13, pp. 11408-11421, Jul., 2022.
\bibitem{b9} X. Chen, J. An, Z. Xiong, C. Xing, N. Zhao, F. R. Yu, and A. Nallanathan, "Covert Communications: A Comprehensive Survey," \emph{IEEE Communi. Surv. \(\&\) Tutori.}, vol. 25, no. 2, pp. 1173-1198, Secondquarter 2023.
\bibitem{b10} M. Shen, K. Ye, X. Liu, L. Zhu, J. Kang, S. Yu, Q. Li, and K. Xu "Machine Learning-Powered Encrypted Network Traffic Analysis: A Comprehensive Survey," \emph{IEEE Communi. Surv. \(\&\) Tutori.}, vol. 25, no. 1, pp. 791-824, Firstquarter 2023.
\bibitem{b11} D. Chen, H. Wang, N. Zhang, X. Nie, H.-N. Dai, K. Zhang, and K.-K. R. Choo, "Privacy-preserving encrypted traffic inspection with symmetric cryptographic techniques in IoT," \emph{IEEE Internet Things J.}, vol. 9, no. 18, pp. 17 265–17 279, sept. 2022.
\bibitem{b12} P. Angueira, I. Val, J. Montalban, O. Seijo, E. Iradier, P. S. Fontaneda, L. Fanari, and A. Arriola, "A survey of physical layer techniques for secure wireless communications in industry," \emph{IEEE Communi. Surv. \(\&\) Tutori.}, vol. 24, no. 2, pp. 810–838, 2022.
\bibitem{b13} S. Han, J. Li, W. Meng, M. Guizani, and S. Sun, “Challenges of physical layer security in a satellite-terrestrial network,” \emph{IEEE Netw.}, vol. 36, no. 3, pp. 98–104, 2022.
\bibitem{b14} Y. Liu, Z. Su, C. Zhang, and H.-H. Chen, “Minimization of secrecy outage probability in reconfigurable intelligent surface-assisted mimome system,” \emph{IEEE Trans. Wireless Commun.}, vol. 22, no. 2, pp. 1374–1387, 2023.
\bibitem{b15} Z. Yin, N. Cheng, T. H. Luan, and P. Wang, “Physical layer security in cybertwin-enabled integrated satellite-terrestrial vehicle networks,” \emph{IEEE Trans. Veh. Technol.}, vol. 71, no. 5, pp. 4561–4572, 2022.
\bibitem{b16} N. Cheng, J. He, Z. Yin, C. Zhou, H. Wu, F. Lyu, H. Zhou, and X. Shen, “6g service-oriented space-air-ground integrated network: A survey,” \emph{Chin. J. Aeronaut.}, vol. 35, no. 9, pp. 1–18, 2022.
\bibitem{b17} Z. Yin, N. Cheng, Y. Hui, W. Wang, L. Zhao, K. Aldubaikhy, and A. Alqasir, “Multi-domain resource multiplexing based secure transmission for satellite-assisted IoT: AO-SCA approach,” \emph{IEEE Trans. Wireless Commun.},  vol. 22, no. 11, pp. 7319-7330, Nov. 2023.
\bibitem{b18} T. K. Rodrigues and N. Kato, “Hybrid centralized and distributed learning for mec-equipped satellite 6g networks,” \emph{IEEE J. Sel. Areas Commun.}, 2023.
\bibitem{b18b} D. Li, Y. Sun, J. Peng, S. Cheng, Z. Yin, N. Cheng, J. Liu, Z. Li, and C. Xu, "Dual Network Computation Offloading Based on DRL for Satellite-Terrestrial Integrated Networks," \emph{IEEE Transactions on Mobile Computing}, vol. 24, no. 3, pp. 2270-2284, March 2025.
\bibitem{b19} W. Chen, X. Lin, J. Lee, A. Toskala, S. Sun, C. F. Chiasserini, and L. Liu, "5G-Advanced Toward 6G: Past, Present, and Future," \emph{IEEE J. Sel. Areas Commun.}, vol. 41, no. 6, pp. 1592-1619, June 2023.
\bibitem{b20} X. Zhu and C. Jiang, “Integrated satellite-terrestrial networks toward 6g: Architectures, applications, and challenges,” \emph{IEEE Internet Things J.}, vol. 9, no. 1, pp. 437–461, 2021.
\bibitem{b21} K. -Y. Lam, S. Mitra, F. Gondesen and X. Yi, "ANT-Centric IoT Security Reference Architecture—Security-by-Design for Satellite-Enabled Smart Cities," \emph{IEEE Internet of Things J.}, vol. 9, no. 8, pp. 5895-5908, 15 April15, 2022.
\bibitem{b22} J. Zhou, H. Pu, H. Cao, C. Cai, P. Guo and H. Jiang, "CORA: Continuous Respiration Monitoring Using Analytical Signal Processing," \emph{IEEE Trans. Mob. Comput.}, vol. 23, no. 12, pp. 13745-13759, Dec. 2024.
\bibitem{b27} A. K. Yadav, "SKAP-NS: A Symmetric Key-Based Authentication Protocol for 5G Network Slicing," \emph{IEEE Trans. Ind. Inf.}, vol. 20, no. 11, pp. 13363-13372, Nov. 2024.
\bibitem{b28} Y. Du, H. Dai, H. Liu, Y. Wang, G. Li, Y. Ren, Y. Chen, and K. Zhang, "Secret Key Generation Based on Manipulated Channel Measurement Matching," \emph{IEEE Trans. Mob. Comput.}, vol. 23, no. 10, pp. 9532-9548, Oct. 2024.
\bibitem{b29} M. Mitev, A. Chorti, H. V. Poor and G. P. Fettweis, "What Physical Layer Security Can Do for 6G Security," \emph{IEEE Open J. Veh. Technol.}, vol. 4, pp. 375-388, 2023.
\bibitem{b30} E. Illi et al., "Physical Layer Security for Authentication, Confidentiality, and Malicious Node Detection: A Paradigm Shift in Securing IoT Networks," \emph{IEEE Commun. Surv. \(\&\) Tutor.}, vol. 26, no. 1, pp. 347-388, Firstquarter 2024.
\bibitem{b31} C. Zhang, J. Yue, L. Jiao, J. Shi and S. Wang, "A Novel Physical Layer Encryption Algorithm for LoRa," \emph{IEEE Commun. Lett.}, vol. 25, no. 8, pp. 2512-2516, Aug. 2021.
\bibitem{b32} X. Lu, J. Lei, Y. Shi and W. Li, "Improved Physical Layer Authentication Scheme Based on Wireless Channel Phase," \emph{IEEE Wireless Commun. Lett.}, vol. 11, no. 1, pp. 198-202, Jan. 2022.
\bibitem{b33} R. Melki, H. N. Noura, M. M. Mansour and A. Chehab, "An Efficient OFDM-Based Encryption Scheme Using a Dynamic Key Approach," \emph{IEEE Internet of Things J.}, vol. 6, no. 1, pp. 361-378, Feb. 2019.
\bibitem{b34} J. Li, J. Wu, L. Chen, J. Li and S. -K. Lam, "Blockchain-Based Secure Key Management for Mobile Edge Computing," \emph{IEEE Trans. Mob. Comput.}, vol. 22, no. 1, pp. 100-114, 1 Jan. 2023.
\bibitem{b35} J . Zhang, A. Marshall, R. Woods and T. Q. Duong, "Design of an OFDM Physical Layer Encryption Scheme," \emph{IEEE Trans. on Veh. Technol.}, vol. 66, no. 3, pp. 2114-2127, Mar. 2017.
\bibitem{b36} L. P. Qian, Y. Wu, N. Yu, D. Wang, F. Jiang and W. Jia, "Energy-Efficient Multi-Access Mobile Edge Computing With Secrecy Provisioning," \emph{IEEE Trans. Mob. Comput.}, vol. 22, no. 1, pp. 237-252, 1 Jan. 2023.
\bibitem{b37} Z. Yin, N. Cheng, T. H. Luan, Y. Hui and W. Wang, "Green Interference Based Symbiotic Security in Integrated Satellite-Terrestrial Communications," \emph{IEEE Trans. Wireless Commun.}, vol. 21, no. 11, pp. 9962-9973, Nov. 2022.
\bibitem{b38} Y. Jiang, Y. Wang, H. Wu, Y. Liu and L. Hu, "Energy-Efficient Covert Offloading in Blockchain-Enabled IoT: Joint Artificial Noise and Computation Resource Allocation," \emph{IEEE Internet of Things J.}, vol. 12, no. 6, pp. 6889-6901, 15 March15, 2025.
\bibitem{b39} Y. Cheng, J. Lu, D. Niyato, B. Lyu, M. Xu and S. Zhu, "Performance Analysis and Power Allocation for Covert Mobile Edge Computing With RIS-Aided NOMA," \emph{IEEE Trans. Mob. Comput.}, vol. 23, no. 5, pp. 4212-4227, May 2024.
\bibitem{b40} R. Chen, M. Liu, Y. Hui, N. Cheng and J. Li, "Reconfigurable Intelligent Surfaces for 6G IoT Wireless Positioning: A Contemporary Survey," \emph{IEEE Internet of Things J.}, vol. 9, no. 23, pp. 23570-23582, Dec. 2022.
\bibitem{b41} Z. Wang, Z. Yin, X. Wang, N. Cheng, Y. Zhang and T. H. Luan, "Label-Free Deep Learning Driven Secure Access Selection in Space-Air-Ground Integrated Networks," \emph{GLOBECOM 2023 - 2023 IEEE Global Communications Conference}, Kuala Lumpur, Malaysia, 2023, pp. 958-963.
\bibitem{b42} Z. Yin, N. Cheng, Y. Song, Y. Hui, Y. Li, T. H. Luan, and S. Yu, "UAV-Assisted Secure Uplink Communications in Satellite-Supported IoT: Secrecy Fairness Approach," \emph{IEEE Internet of Things J.}, vol. 11, no. 4, pp. 6904-6915, 15 Feb, 2024.
\bibitem{b43} P. Liu, J. Si, Z. Li, N. Al-Dhahir and Y. Gao, "Joint 3D trajectory and power optimization for dual-UAV-assisted short-packet covert communications," \emph{IEEE Internet of Things J.}, vol. 11, no. 10, pp. 17388-17401, May 2024.
\bibitem{b44} H. Son and M. Jung, "Phase Shift Design for RIS-Assisted Satellite-Aerial-Terrestrial Integrated Network," \emph{IEEE Trans. on Aerosp. Electron. Syst.}, vol. 59, no. 6, pp. 9799-9806, Dec. 2023. 
\bibitem{b45} K. Yu, Z. Feng, J. Yu, T. Chen, J. Peng and D. Li, "Secure Ultra-Reliable and Low Latency Communication in UAV-Enabled NOMA Wireless Networks," \emph{IEEE Trans. on Veh. Technol.}, vol. 73, no. 10, pp. 14908-14922, Oct. 2024.

\end{thebibliography}
\end{document}